\documentclass[aps,prl,twocolumn,showpacs,floatfix,superscriptaddress]{revtex4}
\usepackage{amsmath,amssymb,amsfonts}
\usepackage{graphics}

\begin{document}

\title{EXAFS, XRD and RMC studies of an 
Amorphous Ga$_{50}$Se$_{50}$ Alloy Produced by Mechanical Alloying}

\author{K. D. Machado}
\email{kleber@fisica.ufsc.br}
\affiliation{Departamento de F\'{\i}sica, Universidade Federal de Santa Catarina, 88040-900 
Florian\'opolis, SC, Brazil}

\author{P. J\'ov\'ari}
\email{pjovari@mail.desy.de}
\affiliation{Hamburger Synchrotronstrahlungslabor HASYLAB am Deutschen
Elektronen-Synchrotron DESY, Notkestrasse, 85, D-22603, Hamburg, Germany}

\author{J. C. de Lima}
\affiliation{Departamento de F\'{\i}sica, Universidade Federal de Santa Catarina, 88040-900 
Florian\'opolis, SC, Brazil}

\author{C. E. M. Campos}
\affiliation{Departamento de F\'{\i}sica, Universidade Federal de Santa Catarina, 88040-900 
Florian\'opolis, SC, Brazil}

\author{T. A. Grandi}
\affiliation{Departamento de F\'{\i}sica, Universidade Federal de Santa Catarina, 88040-900 
Florian\'opolis, SC, Brazil}

\date{\today}

\begin{abstract}

The local atomic order of an amorphous Ga$_{50}$Se$_{50}$ alloy produced by Mechanical Alloying 
(MA) 
was studied by the Extended X-ray Absorption Fine Structure (EXAFS) and X-ray Diffraction (XRD) 
techniques and by Reverse 
Monte Carlo (RMC) 
simulations of its total x-ray structure factor. The coordination numbers and interatomic distances 
for the first neighbors were determined by means of EXAFS analysis and RMC simulations. The RMC 
simulations also furnished 
the partial pair distribution functions $G^{\text{RMC}}_{\text{Ga-Ga}}(r)$, 
$G^{\text{RMC}}_{\text{Ga-Se}}(r)$ and $G^{\text{RMC}}_{\text{Se-Se}}(r)$. The 
results obtained indicated that there 
are important differences among the local structure of the amorphous Ga$_{50}$Se$_{50}$ 
alloy produced by MA and those of the corresponding crystals, since there are Se-Se pairs 
in the first coordination shell of the amorphous alloy that are forbidden in the Ga$_{50}$Se$_{50}$ 
crystals.

\end{abstract}


\maketitle

In the recent years there has been an increase in the number of applications related to nonlinear 
optical materials. However, the desired properties concerning this kind of applications, such as 
optical homogeneity, laser damage threshold, stability of the compound upon exposure to laser 
beam, ease of fabrication, improved mechanical strength and the possibility of making large 
crystals are difficult to find in a single material. Some of the materials that can be used for 
nonlinear applications, like silver gallium selenides, zinc germanium phosphides and thallium 
arsenic selenides, do not fulfill all of these requirements, limiting severely their efficiency 
and applicability. Thus, there is a high necessity of developing new materials with a higher level 
of performance and more cost effective characteristics. Gallium selenide (GaSe) has a number of 
interesting properties for electrical and nonlinear optics applications. It transmits in the 
wavelength range varying from 0.65 to 18 $\mu$m and its optical absorption coefficient remains 
below 1 cm$^{-1}$ throughout the transparency range. It has the possibility of converting sum and 
difference frequencies \cite{fernelius,singh2}. It has also been reported to be used in making 
a number of devices like MOSFET, IR detector, Solar Cell, compound semiconductor, etc. in crystalline 
form while in amorphous form, it is a potential candidate for optical memory type 
applications \cite{singh,singh2}. Crystalline GaSe is a semiconductor of the III-VI family like 
GaS and InSe and it has a layered graphite type structure with a fourfold layer in the sequence 
Se-Ga-Ga-Se. The crystal cleaves very easily along the layers \cite{fernelius}. At room 
temperature, the layers are bound by weak van der Waals-type interactions. The weakness of this 
interaction explains the existence of a number of polytypes \cite{jandl}. 
Ga$_{50}$Se$_{50}$ alloys can be prepared by the melting, vapor deposition and molecular beam 
epitaxy techniques \cite{Ludviksson,Fujita,Stoll,Ng}. These techniques have had very limited success 
because they do not have control over the kinetics and morphology. In addition, due to the low melting 
points of the elemental Ga (30$^\circ$C) and Se (217$^\circ$C) and the high vapor pressure of Se 
above 600$^\circ$C it is difficult to obtain Ga-Se alloys at specific compositions. On the other 
hand, the mechanical alloying (MA) technique \cite{Mec} can be used to overcome these difficulties 
since the temperatures reached in MA are very low, what reduces reaction kinetics, allowing the 
production of poorly crystallized or amorphous materials \cite{kleber,carlos1,carlos2,Lima3} even 
if the constituents of the alloy have low melting points, as it is in the case of gallium and 
selenium.

\begin{figure}
\includegraphics{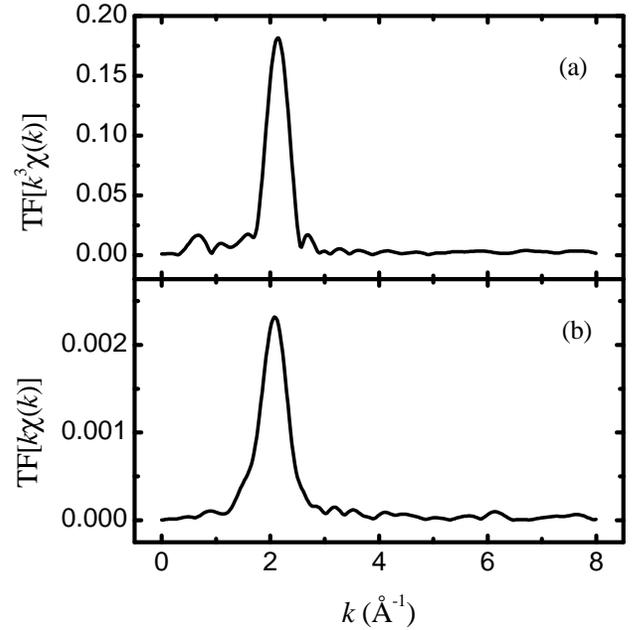}
\caption{\label{fig2} Fourier transformation of experimental EXAFS spectra: a) at the 
Ga K-edge and b) at the Se K-edge.}
\end{figure}

\begin{figure}
\includegraphics{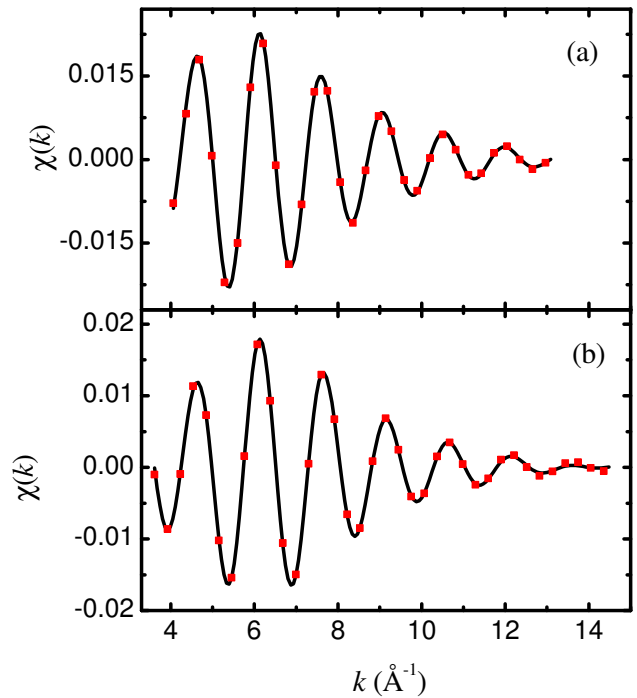}
\caption{\label{fig34} Fourier-filtered first shell (full line) and its simulation (squares) for 
{\em a}-Ga$_{50}$Se$_{50}$ at the (a) Ga K edge, (b) Se K edge.}
\end{figure}

Amorphous Ga$_{50}$Se$_{50}$ ({\em a}-Ga$_{50}$Se$_{50}$) was prepared by MA starting from high 
purity elemental powder selenium (Alfa Aesar, 99.9999\% purity, 150 $\mu$m) and scraped ingots of 
gallium (Aldrich, 99.999\% purity) with nominal composition Ga$_{50}$Se$_{50}$. The mixture was 
sealed together with several steel balls into a cylindrical steel vial under argon atmosphere. 
The ball-to-powder weight ratio was 10:1. A Spex Mixer/Mill model 8000 was used to perform MA at 
room temperature. The mixture was milled for 15 h. A ventilation system was used to keep the vial 
temperature close to room temperature. The composition of the alloy was confirmed by an energy 
dispersive spectroscopy (EDS) measurement and impurity traces were not observed. The alloy produced 
was investigated by extended x-ray absorption fine structure (EXAFS) and x-ray diffraction (XRD) 
techniques and also by reverse Monte Carlo simulation (RMC). The EXAFS measurements were carried 
out on the D04B beam line of LNLS (Campinas, Brazil), using a channel cut monochromator (Si 111), 
two ionization chambers as detectors and a 1 mm entrance slit. All data were taken at room 
temperature in the transmission mode. The EXAFS oscillations $\chi(k)$ at both K edges, after the 
standard data reduction procedures using Winxas97 software \cite{exafs2}, were Fourier 
transformed (FT) using a Hanning weighting function within the ranges 3.8 -- 14.3 \AA$^{-1}$ for 
the Ga, and 3.4 -- 14.3 \AA$^{-1}$ for the Se edge. They can be seen in Fig.~\ref{fig2}. Raw spectra 
were filtered by Fourier transforming $k^3\chi(k)$ (Ga edge) and $k\chi(k)$ (Se edge) into $r$-space 
(Fig.~\ref{fig2}) and transforming back the first coordination shells (1.20 -- 2.85 \AA\ for the Ga 
edge and 1.13 -- 3.0 \AA\ for the Se edge). Filtered spectra were then fit by using Gaussian 
distributions to represent the homopolar and heteropolar bonds \cite{Stern}. The amplitude and 
phase shifts relative to the homopolar and heteropolar bonds needed to fit them were obtained from 
{\em ab initio} calculations using the spherical waves method \cite{Rehr} and by the FEFF software. 
Figure~\ref{fig34} shows the experimental and the fitting results for the Fourier-filtered first 
shells on the Ga and Se edges. Structural parameters extracted from the fits are listed in 
Table \ref{tabI}. It is interesting to note that the very good fits shown in Fig.~\ref{fig34} were 
achieved only when Se-Se pairs were considered in the first shell. This fact indicates that the 
local structure of {\em a}-Ga$_{50}$Se$_{50}$ produced by MA is different from its crystalline 
counterparts as none of the known stable crystalline Ga$_{50}$Se$_{50}$ structures contains 
Se-Se bonds. 

\begin{figure}[h]
\includegraphics{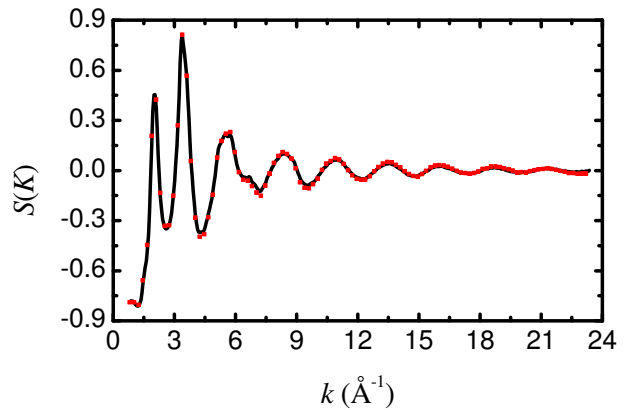}
\caption{\label{fig5} Experimental (full line) and simulated 
(squares) total structure factor for {\em a}-Ga$_{50}$Se$_{50}$.}
\end{figure}

The XRD measurements were carried out at the BW5 beamline \cite{Palref} at HASYLAB. All data were 
taken at room temperature. The energy of the incident beam was 121.3 keV ($\lambda=0.102$ \AA). The 
structure factor ${\cal S}(K)$ (Fig.~\ref{fig5}, full line) was computed from the normalized 
intensity $I(K)$ according to Faber and Ziman \cite{Faber} and it was modeled by reverse Monte Carlo 
simulations. This technique is described in details elsewhere \cite{RMC1,RMCA,rmcreview,kleber} and 
it has been used as a method for structural modeling based directly on experimental data. There are 
several papers \cite{RMC6,kleber,RMC7,RMC9,RMC11,RMC13} reporting structural studies of amorphous 
alloys by RMC. Simulations were carried out by the RMC program available on the Internet 
\cite{RMCA}. Cubic cells contained 1600 and 12800 atoms, and the average density was 
$\rho_0 = 0.03907 \pm 0.0005$ atoms/\AA$^3$. This value was found from the slope of the straight 
line ($-4\pi\rho_0 r$) fitting the initial part (until the first minimum) of the total 
$G(r)$ function \cite{Waseda}. 
The minimum distance of atoms was also extracted from $G(r)$ and fixed at 2.18 \AA. All the 
simulations were performed considering atoms randomly placed in the cubic cells as starting 
configurations. Then the following series of simulations were carried out: 

\begin{figure}
\includegraphics{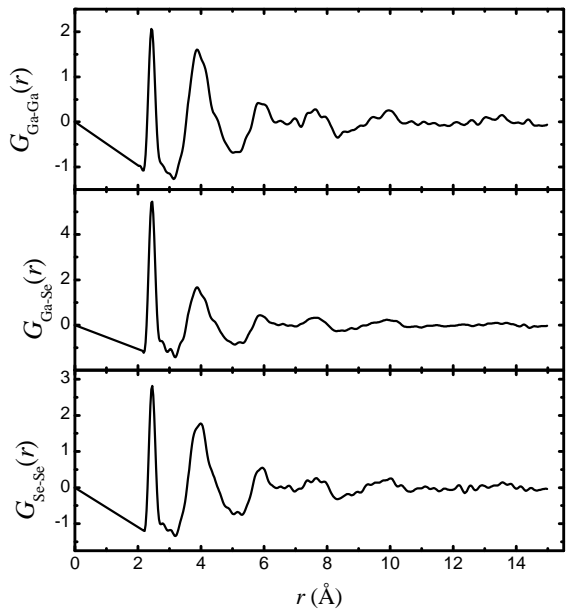}
\caption{\label{fig7} $G_{\text{Ga-Ga}}^{\text{RMC}}(r)$, 
$G_{\text{Ga-Se}}^{\text{RMC}}(r)$ and $G_{\text{Se-Se}}^{\text{RMC}}(r)$ 
functions obtained from the RMC simulations.}
\end{figure}

\begin{enumerate}
\item Hard sphere simulation without experimental data to avoid possible memory effects of the 
initial configurations in the results. 

\item `Unconstrained' runs (i.e. when experimental data were `switched on'). These runs led to three 
essentially identical partial pair correlation functions and partial structure factors which can be 
considered as linear combinations of the `true' partial quantities. It is to be mentioned that as 
neither the size nor other {\em a priori} information can distinguish between Ga and Se atoms at this 
step no adequate coordination numbers can be obtained.

\item `Constrained' runs. The experimental ${\cal S}(K)$ was fit by using EXAFS coordination number 
values as constraints. Comparison of experimental (full line) and calculated (squares) structure 
factors for the latter case is shown in Fig.~\ref{fig5} and the partial pair correlation 
functions (pcf's) are given in Fig.~\ref{fig7}. 
\end{enumerate}

\begingroup
%\squeezetable
\begin{table}
\caption{\label{tabI} Structural parameters obtained for {\em a}-Ga$_{50}$Se$_{50}$.}
\begin{ruledtabular}
\begin{tabular}{ccccc}
\multicolumn{5}{c}{}\\[-0.4cm]
\multicolumn{5}{c}{EXAFS} \\[0.1cm]\hline\hline
&  &  && \\[-0.3cm]
& \multicolumn{2}{c}{Ga K-edge} & \multicolumn{2}{c}{Se K-edge}\\[0.1cm]\hline
& & & & \\[-0.3cm]
Bond Type & Ga-Ga & Ga-Se & Se-Ga & Se-Se\\[0.1cm]
$N$ & 1.3  & 2.4  & 2.4  & 1.3  \\
$r$ (\AA) & 2.38 & 2.45  & 2.45  & 2.37\\
$\sigma^2$ (\AA $\times 10^{-2}$) & 1.58  & 0.545 & 0.545 & 1.77 \\\hline\hline
\multicolumn{5}{c}{}\\[-0.3cm]
\multicolumn{5}{c}{RMC} \\[0.1cm]\hline\hline
& & & & \\[-0.3cm]
Bond Type & Ga-Ga & Ga-Se & Se-Ga & Se-Se\\[0.1cm]
$N$ & 1.2 & 2.5 & 2.5 & 1.3 \\
$r$ (\AA) & 2.42 & 2.42 & 2.42 & 2.42\\\hline\hline
\multicolumn{5}{c}{}\\[-0.3cm]
\multicolumn{5}{c}{Ga$_{50}$Se$_{50}$ compound \footnote{Space group P63/MMC.}} \\[0.1cm]\hline\hline
& &  &  & \\[-0.3cm]
Bond Type & Ga-Ga & Ga-Se & Se-Ga & Se-Se\\[0.1cm]
$N$ & 1 & 3 & 3 & 6 \\
$r$ (\AA) & 2.44 & 2.45 & 2.45 &3.75\\\hline\hline
\multicolumn{5}{c}{}\\[-0.3cm]
\multicolumn{5}{c}{Ga$_{50}$Se$_{50}$ compound \footnote{Space group P-6M2}} \\[0.1cm]\hline\hline
& &  &  & \\[-0.3cm]
Bond Type & Ga-Ga & Ga-Se & Se-Ga & Se-Se\\[0.1cm]
$N$ & 1 & 3 & 3 & 6 \footnote{The trigonal crystal of space group R3M has 4 Se-Se pairs.} \\
$r$ (\AA) & 2.39 & 2.47 & 2.47 & 3.74
\end{tabular}
\end{ruledtabular}
\end{table}
\endgroup

Finally the whole series of calculations was repeated from the very beginning with the difference 
that during the `constrained' run random steps resulting in non-zero Se-Se first coordination number 
were rejected. It is important to note that if Se-Se pairs are forbidden as first neighbors 
simulations did not converge, reinforcing the results obtained by EXAFS analysis. The position of 
the first and second peak are 2.42 \AA\ and 3.89 \AA\ in all of the pcf's corresponding to a mean 
bond angle of 107$^\circ$ for the four bond types (Ga-Ga-Ga, Se-Se-Se, Ga-Se-Se, Ga-Ga-Se) that can 
be directly derived from the pcf peak positions. As this is very close to the value describing perfect 
tetrahedral coordination (109.5$^\circ$) and $N_{\text{Ga-Ga}} + N_{\text{Ga-Se}}$ and 
$N_{\text{Se-Se}} + N_{\text{Se-Ga}}$ are both close to 4 it is evident to assume that ball milled 
{\em a}-Ga$_{50}$Se$_{50}$ has a tetrahedral structure with a definite tendency to form homopolar 
bonds.

The difference of Ga-Ga and Ga-Se bond lengths in the crystalline modifications is not greater than 
about 0.08~\AA\ (see Table~\ref{tabI}) and they are also quite close to the value of 2.35 \AA\ 
found recently for {\em a}-Se \cite{RMC6}. As the spatial resolution of diffraction experiments is 
equal to $\pi/K_{\text{max}}$ the ${\cal S}(K)$ factor should be measured at least up to 
40~\AA$^{-1}$ or further to get more detailed information on the first coordination shell. It 
should also be mentioned that due to the value of neutron scattering lenghts 
($b_{\text{Se}}=7.970$ fm, $b_{\text{Ga}}=7.288$ fm) neutron diffraction data would give essentially 
the same information. Other techniques used to obtain information at the level of pcf's are either 
prohibitively expensive (neutron diffraction with isotopic substitution) or yield limited spatial 
resolution due to the low $K_{\text{max}}$ value available (anomalous X-ray scattering).

In summary the local structure of ball milled {\em a}-Ga$_{50}$Se$_{50}$ was investigated 
experimentally with EXAFS and high energy X-ray diffraction. EXAFS analysis led to the following 
conclusions: the average first coordination number in {\em a}-Ga$_{50}$Se$_{50}$ is close to 4, Ga 
and Se local environments are similar and Se-Se bonding is significant. All of these findings were 
checked and confirmed by RMC study of diffraction data: it was possible to obtain a good fit with 
coordination constraints close to the EXAFS values while runs without Se-Se first neighbors led to 
a bad agreement between model and experiment. Mean bond angle calculated from diffraction data is 
107$^\circ$ indicating that {\em a}-Ga$_{50}$Se$_{50}$ has a tetrahedral local structure.

The present study illustrates how complementary information obtained by different experimental 
techniques can be combined within the frame of reverse Monte Carlo simulation. We believe that this 
is a useful and efficient way of modelling disordered materials especially in cases when traditional 
methods (e.g. neutron diffraction with isotopic substitution) are not available.

\acknowledgments

The Brazilian authors wish to thank the Brazilian agencies CNPq and CAPES for financial 
support. This study was also partially supported by LNLS (Proposal No. XAS 998/01). P. J\'ov\'ari 
is indebted to Hermann Franz and Martin von Zimmermann (both HASYLAB) for their help during the 
diffraction measurement.

%\bibliographystyle{apsrev}
%\bibliography{ga50se50}

\end{document}